\documentclass[sigconf]{acmart}

\usepackage{kotex} \usepackage{tabularx}
\usepackage{multirow}
\usepackage{todonotes}
\usepackage{graphicx}
\usepackage{subfig}

\AtBeginDocument{}

\setcopyright{acmlicensed}
\copyrightyear{2025}
\acmYear{2025}
\acmDOI{XXXXXXX.XXXXXXX}

\acmConference[XX]{}{June 03--05,
  2018}{Woodstock, NY}
\acmISBN{978-1-4503-XXXX-X/18/06}

\begin{document}

\title{Beyond Turn-taking: Introducing Text-based Overlap into Human-LLM Interactions}

\author{JiWoo Kim}
\email{wldn9705@skku.edu}
\affiliation{
  \institution{Sungkyunkwan University}
  \city{Suwon}
  \country{South Korea}
  }

\author{Minsuk Chang}
\authornote{Corresponding authors}
\email{minsukchang@google.com}
\affiliation{% 
\institution{Google DeepMind}
  \city{Seattle}
  \country{USA}
}

\author{JinYeong Bak}
\authornotemark[1]
\email{jy.bak@skku.edu}
\affiliation{
  \institution{Sungkyunkwan University}
  \city{Suwon}
  \country{South Korea}
  }

\renewcommand{\shortauthors}{Kim et al.}

\begin{abstract}
Traditional text-based Human-AI interactions often adhere to a strict turn-taking approach. In this research, we propose a novel approach that incorporates overlapping messages, mirroring natural human conversations. Through a formative study, we observed that even in text-based contexts, users instinctively engage in overlapping behaviors like ``A: Today I went to--'' ``B: yeah.'' To capitalize on these insights, we developed OverlapBot, a prototype chatbot where both AI and users can initiate overlapping. Our user study revealed that OverlapBot was perceived as more communicative and immersive than traditional turn-taking chatbot, fostering faster and more natural interactions. Our findings contribute to the understanding of design space for overlapping interactions. We also provide recommendations for implementing overlap-capable AI interactions to enhance the fluidity and engagement of text-based conversations.

 \end{abstract}




\begin{teaserfigure}
    \centering
  \includegraphics[width=\textwidth]{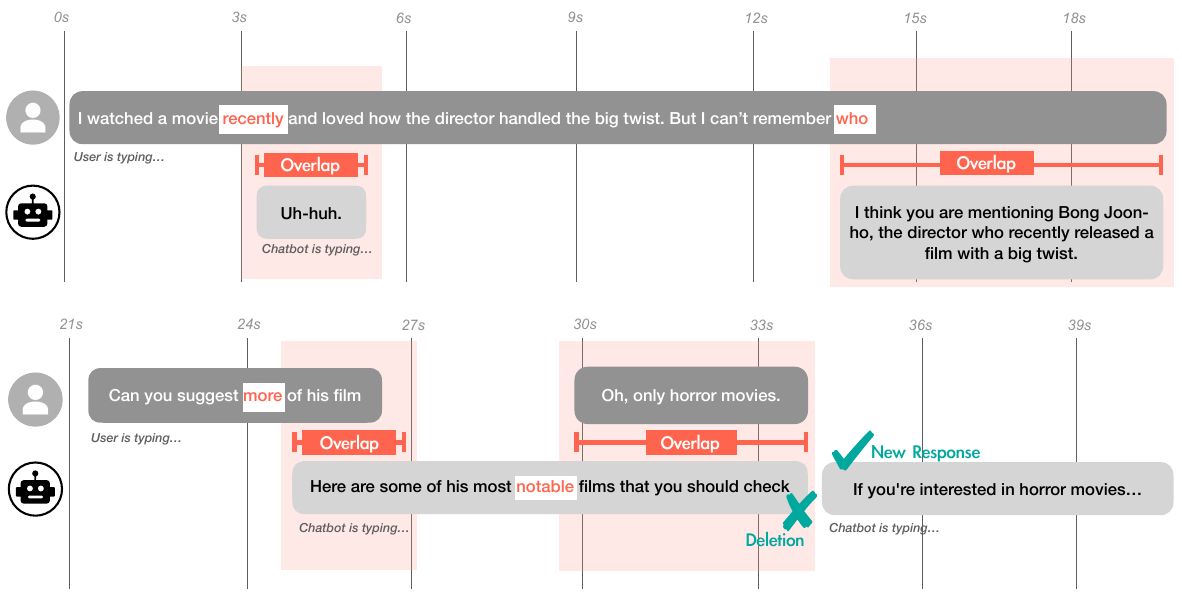}
  \caption{Introduction of overlapping into text-based chat interaction with AI: The chatbot generates overlapping messages with the user's typing through backchanneling and preemptive answers before the user finishes typing. The user overlaps with the chatbot's typing by interrupting it. In such cases, the chatbot deletes its previous response and regenerates a new answer based on the interruption.}
\label{fig:systemfigure}
\end{teaserfigure} 
\received{20 February 2007}
\received[revised]{12 March 2009}
\received[accepted]{5 June 2009}

\maketitle

\section{Introduction}
\label{introduction}
Human-to-human conversations differ from chess, where turns are strictly alternated. In human-to-human conversation, overlaps and interruptions are common, requiring participants to coordinate who speaks, when to stop, and when to continue \cite{duncan1972some}. Humans are adept at this, often achieving smooth interactions with minimal gaps, and occasional overlaps that contribute to the fluent flow of the interaction \cite{sacks1974simplest}. On the other hand, current text-based chat interactions follow strict turn-taking, similar to playing chess. This applies not only to human-human interactions but also to interactions with Large Language Models (LLM), where users must wait for the chatbot to respond before the conversation can be continued (Figure \ref{fig:strict-turn-taking}) \cite{DBLP:journals/chb/ZhouLHJ23}. 

In light of this, we have identified a new opportunity for designing \textit{text-based overlap} in \textit{human-LLM} interactions. The most closely related prior research focuses on real-time messaging \textit{between humans}. These studies, where individuals can see each other’s typing \cite{podlubny2017synchronous, iftikhar2023together, DBLP:conf/chi/Phillips00} and simultaneously send overlapping messages \cite{kim2017applying, DBLP:conf/uist/VronaySD99, DBLP:conf/chi/Phillips00}, have mainly focused on the effects of these interactions such as minimizing pauses and enhancing collaboration. By investigating the specific overlapping behaviors and their underlying motivations in detail, we can deepen our understanding of how people interact in overlap-capable text-based interactions and enable the chatbot to handle overlaps more seamlessly. This need for further exploration highlights the importance of additional investigation, which leads us to the following question:

\begin{description}
    \item[RQ1] What actionable behaviors can an LLM take to overlap with a user’s typing?
\end{description}

To investigate RQ1, we conducted a formative study with a research probe \cite{DBLP:conf/chi/WallaceMWO13} that allowed seven pairs of participants to overlap with their interlocutors by typing concurrently. The probe allowed the participants to see each other's messages while they were still being written on the screen and to send messages even while the other person was typing. We observed that participants instinctively engage in text-based overlapping interactions. Specifically, they overlapped their typing with the interlocutor's typing by 1) giving preemptive answers to the interlocutor's questions and 2) providing brief listening cues or instant agreement, i.e., backchanneling. They also decided who should type next by 3) deleting their typed messages, indicating an overlap-capable interface provides a new channel for participants to express their messages \cite{kim2017applying}. 

Grounded on these observations, we hypothesized these traits could enhance the naturalness and efficacy of human-LLM interactions (Figure \ref{fig:systemfigure}). (a) By facilitating real-time typing, akin to the simultaneous listening and comprehension that characterizes human conversations \cite{skantze2021turn}, and (b) by providing immediate feedback, including active listening cues \cite{heinz2003backchannel} or quick responses \cite{schegloff2000overlapping}, we sought to emulate the dynamic nature of human discourse. To comprehensively explore the design space for overlapping interactions, we formulated the following research questions: 

\begin{figure*}
    \centering
    \includegraphics[width=1\textwidth]{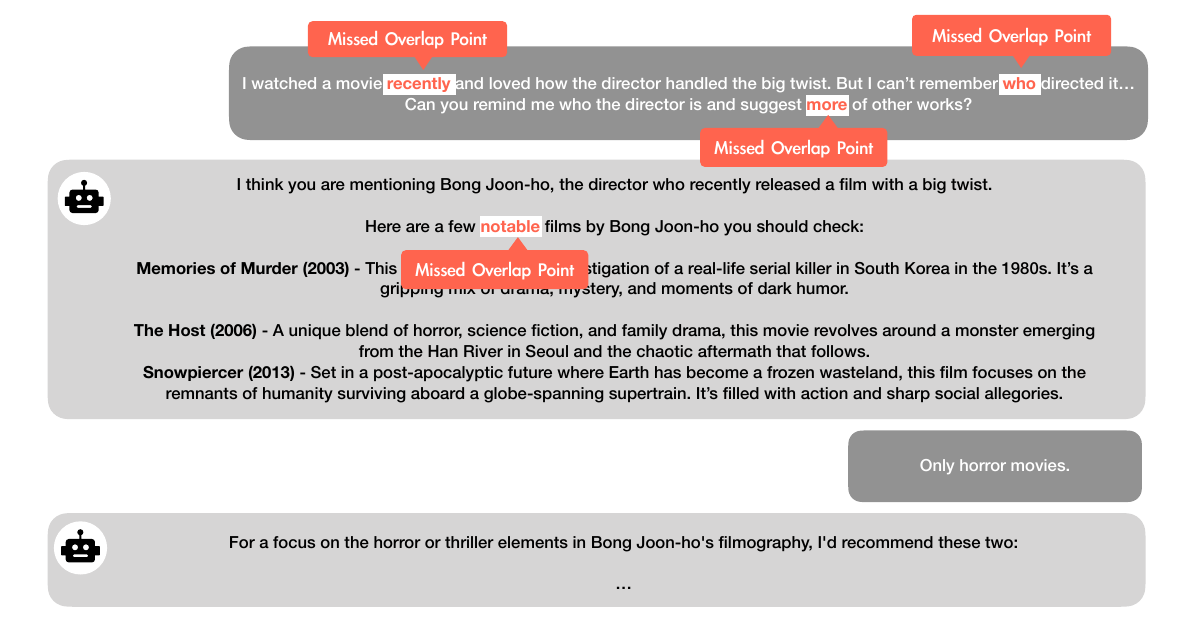}
    \caption{Current turn-taking chat with an LLM. Motivated by the absence of overlap points, we identified a new opportunity to design overlap-capable text-based interactions.}
    \label{fig:strict-turn-taking}
\end{figure*} 
\begin{description}
    \item[RQ2] How can we enable an LLM to replicate similar overlapping behaviors?
    \item[RQ3] How will users perceive overlapping interactions with AI, and what new interaction patterns might develop?
    \item[RQ4] What challenges could arise, and what key considerations should guide the design of this feature?
\end{description}

We developed OverlapBot, an LLM-powered AI chatbot which overlaps its typing with user's typing by 1) giving preemptive responses and 2) making backchanneling, sometimes 3) deleting its own typing. To achieve this, we finetuned an open-source LLM with publicly available datasets engineered to our purpose. In a within-subject user study with 18 participants engaged in free-topic conversations with OverlapBot, we found that OverlapBot was perceived as more communicative and immersive than traditional turn-taking chatbot, facilitating speedy interactions. We analyzed interview data to provide insights for designing AI that effectively manages overlap in text-based interactions.

We identified three new text-based human-LLM interactions that resembled human-to-human conversations, differing from traditional turn-taking chatbots. First, participants checked the chatbot's attentiveness from its preemptive responses, similar to how they use active listening signals in human interactions \cite{heinz2003backchannel}. Second, participants did not respond verbally to the chatbot’s backchanneling but acknowledged its presence, much like they would in human interactions. Third, participants issued short interruption commands to stop the chatbot from typing, just as they would interrupt someone to stop them from speaking. These findings suggest that people instinctively engage in overlapping behaviors during text-based human-LLM interactions, a dynamic traditional turn-taking chatbots lack. When allowed to overlap, people instinctively embrace it, revealing new possibilities for chatbot design.

In summary, our contributions include: 
\begin{itemize}
\item Introduction and exploration of design space for overlapping interactions in text-based communication.
\item Development of OverlapBot, an interactive AI chat prototype that supports overlapping and interrupting. 
\item User study findings demonstrating that users find text-based overlap in human-LLM interactions to be natural and efficient.
\item A set of design insights for implementing overlap-able interactions in AI systems.
\end{itemize}

\section{Background}
\label{related-work}

\subsection{Overlap in Human Communication, Cooperative Overlap and Competitive Overlap}
\label{related-work-overlap}
Human-to-human conversations generally follow a pattern where one person speaks at a time, yet overlap in speech is a frequent occurrence \cite{skantze2021turn, zimmermann1996sex}. It is important to recognize that these overlaps should not merely be viewed as failures in turn-taking, as they often fulfill important functions and contribute to the smooth flow of interaction \cite{coates1994no}. Overlapping speech is not always a sign of dominance or unfriendliness \cite{goldberg1990interrupting}. Previous studies have identified two distinct types of overlap: cooperative and competitive \cite{schegloff2000overlapping, murata1994intrusive, DBLP:journals/taffco/EgorowW22}. Cooperative overlap involves both speakers contributing to the conversation collaboratively, without competing for control. A common example of this is back-channeling \cite{DBLP:conf/icls/Yardi06, heinz2003backchannel}, where the listener provides brief, often subtle vocalizations such as ``mm hmm,'' ``uh huh,'' or ``yeah.'' These responses, although frequent, are not typically considered full ``turns'' in conversation. Another form of cooperative overlap is terminal overlap, where the listener anticipates the speaker’s turn ending and begins to speak before the turn is fully completed. Conversely, competitive overlap occurs when speakers vie for control of the conversation, with one eventually needing to relinquish their turn. Unlike cooperative overlap, competitive overlap requires a resolution mechanism to determine which speaker should continue \cite{goldberg1990interrupting, skantze2021turn}. Previous research highlights that while overlaps can be objectively identified in a corpus, interruptions require interpretation, as one speaker is seen as violating the other's right to speak \cite{bennett1978interruptions}.

\subsection{Refining turn-taking interaction designs in Speech Conversation and Robotics}
Previous studies, by refining turn-taking human-AI systems, have sought to facilitate more natural dialogues that resemble human-to-human interactions \cite{DBLP:conf/cui/AylettR23, DBLP:conf/hai/AylettSMSRFR23, DBLP:conf/iva/EhretBNEMSFK23, DBLP:conf/iva/JanowskiA18, DBLP:journals/csl/Skantze21}. In particular, research into speech-based interaction systems \cite{DBLP:conf/inhci/PhukonSB22, ma2024languagemodellistenspeaking, DBLP:conf/interspeech/ChangLSZSLH22} has concentrated on minimizing the awkward silences during exchanges, addressing the crucial aspect of timing in conversations. Various methodologies, including silence-based, inter-pausal unit-based, and continuous turn-taking models, have been developed to modulate these behaviors. These models utilize verbal cues such as pitch and speech rate to finetune interaction timings. In addition, research into robotic systems \cite{DBLP:conf/hri/Paetzel-Prusmann23, DBLP:conf/icmi/LalaIK19, DBLP:conf/hri/MoujahidHL22} have demonstrated that enhancing turn-taking capabilities contributes significantly to the naturalness of multiparty conversations. These studies have explored classifiers for detecting turn-holding cues, incorporating analyses of various auditory and visual signals, including volume, pitch, speech rate, and gaze. This emphasis extends beyond verbal cues to encompass nonverbal elements such as gaze and gestures, underlining their importance in conversational dynamics. Conversely, text-based interactions, which rely solely on semantic cues, represent a relatively underexplored area in refining turn-taking strategies. Our work shares a common motivation with \cite{DBLP:journals/corr/abs-2406-15718}, who introduced duplex models capable of generating responses concurrently with input reception. In contrast to their work, our research emphasizes the importance of HCI findings for interaction design and offers insights into incorporating natural conversational behavior in text-based human-LLM interactions.

\subsection{Real-time Text Messaging}
\label{related-work-realtime-message}
Research on real-time messaging in text-based interaction has uncovered various effects on collaboration and communication \cite{rejhon2013standardization, iftikhar2023together}. Some studies have shown that when messages are visible to interlocutors as they are being typed, user coordination improves and message editing decreases \cite{solomon2010speaking, dringus1991study}. Field trials have indicated that synchronous communication can foster greater cooperation and engagement, particularly in close relationships \cite{podlubny2017synchronous}. Further studies have suggested that real-time messaging enhances conversational experiences by minimizing silence and incorporating nonverbal cues, such as pauses and typing speed, into the communication process \cite{kim2017applying}. These findings illustrate the positive impact of real-time messaging, highlighting its potential to facilitate smoother interactions. Our study differs by enabling real-time text-based messaging between a human and an LLM-powered chatbot, where the chatbot is inherently capable of managing overlap. 

\subsection{Large Language Models, Text-based Conversational Agent, Interactive Designs}
Recent advancements have led to the widespread development of Large Language Models, or text-based conversational agents (LLMs). LLMs are increasingly being applied across various domains due to their interactivity \cite{min2023recent, shahriar2023let, dang2022prompt, white2023prompt, park2023generative}. These interactions typically rely on verbose textual prompting, sometimes complemented by graphical manipulations such as buttons or mouse pointer movements.

\subsubsection{Verbose Textual Prompting}
The primary mode of interaction with LLMs is through a prompting interface \cite{DBLP:journals/tist/ChangWWWYZCYWWYZCYYX24}. Users craft specific prompts to guide LLMs in performing tasks such as email generation, text summarization, or question-answering. Additionally, users can engage in dialogue-like interactions, allowing for natural language conversations with the models.
Several widely adopted techniques enhance textual prompting. For instance, the Chain-of-Thought method enables LLMs to provide step-by-step reasoning \cite{DBLP:conf/acl/0009C23, DBLP:conf/nips/Wei0SBIXCLZ22}, while Multi-Turn instructions allow for iterative problem-solving by incorporating user feedback into subsequent prompts \cite{DBLP:journals/corr/abs-2307-06435}. These approaches align with a strict turn-taking conversational paradigm, where users input a prompt, wait for the model’s response, and repeat the process. However, few studies have explored interaction paradigms that move beyond traditional turn-taking in text-based human-AI exchanges.
Our work introduces overlapping capabilities to LLMs, broadening the interaction design space by enabling overlapping functionality. This enables forms of interaction that expands the possibilities for ``how'' users and LLM can interact with.

\subsubsection{Graphical Manipulations Combined with Textual Prompting}
Many integrations of LLMs incorporate graphical elements \cite{DBLP:conf/uist/JiangRDX23, DBLP:conf/uist/SuhMPX23}, including widgets like buttons and sliders to trigger predefined textual or system prompts. For example, buttons are often used as shortcuts for tasks such as editing text or generating code \cite{DBLP:conf/iui/YuanCRI22, DBLP:conf/iui/ClarkRTJS18}. OpenAI’s ChatGPT API, for instance, includes a ``stop generating'' button, which requires users to use their mouse to pause the model's response. In comparison, our proposed interface enables users to stop the chatbot by simply sending a textual command that overlaps with the ongoing interaction. In addition, sliders are commonly utilized to adjust model parameters, allowing users to modify continuous variables that affect the generation of outputs like images or music \cite{DBLP:conf/chi/DangMB22, DBLP:conf/chi/LouieCHTC20}. In addition, gestures and physical metaphors are sometimes employed to refine LLM outputs. For example, pointing to a specific area can highlight elements of an image or guide the model to regenerate only a selected part \cite{DBLP:journals/corr/abs-2305-05662}. Similarly, dragging gestures can be used to adjust spatial attributes of an image, such as pose, facial expressions, or layout \cite{DBLP:conf/chi/MassonMC024, DBLP:conf/siggraph/PanTLLMT23}. Our proposed interface eliminates the need for buttons, sliders, or gestures. Instead, it relies exclusively on text-based interactions, such as stopping the LLM’s response by overlapping functionality.

\section{Study 1 - Exploring Actionable Behaviors for Text-based Overlap}
\label{formative-study}
To explore actionable behaviors that can be performed by LLM in text-based interactions, we conducted a formative study where we built a real-time chat web interface as a research probe. With this probe, users could see each other's typing and could type simultaneously when their interlocutor is typing. This research probe served as a ``tool for design and understanding,'' \cite{DBLP:conf/chi/WallaceMWO13} rather than a prototype interface meant to introduce new interaction techniques. We used it as a means to observe and compare user behavior in relation to different types of overlap, drawing parallels with those observed in speech-based interactions. 

In this experiment, we focused on a task that could induce users to naturally overlap with each other in text-based interaction. As chat conversations can vary depending on the relationship between the partners, we gathered participants by purposive sampling \cite{tongco2007purposive}. A total of 14 participants took part in discussions using our research probe. Their average age was 26 (SD = 2.09), and 8 of them were female and 6 male. 12 participants were native South Korean speakers, 1 participant was a native German speaker, and 1 participant was a native Chinese speaker. These participants formed seven pairs, with six pairs conversing in Korean and one pair (German and Chinese) using English. The pairs were intentionally made up of individuals with different levels of familiarity, including close friends, colleagues, and strangers. 

To encourage conversation, we instructed the pairs to decide on things about a group retreat workshop. They had to decide on three songs to listen to, three dinner menus, and three movies to watch. They were given a 10-minute time limit for these decisions. After the discussion, participants were asked to complete open-ended questions about their overall experience and their intention to use it in the future. We interviewed them when more detailed explanations were needed in open-ended responses. All conversations were recorded, and the participants’ typing logs were saved as files, with their consent.

We collected three types of data: open-ended survey responses, interview transcripts and recorded videos. By observing the recorded videos, we were able to determine the types of overlapping behaviors that occurred. By having the first author and an independent researcher thematically analyze the open-ended responses and interview transcripts \cite{boyatzis1998transforming}, we were able to understand the intentions behind the overlapping behaviors.

\subsection{Findings}
First, we observed that participants frequently engaged in overlapping behaviors. Specifically, participants overlapped with their interlocutor’s typing by starting to type even before the other person finished typing. All participants showed and acknowledged this behavior. Participants reported their intentions as follows, which were related to cooperative overlap (Section \ref{related-work-overlap}).

\begin{enumerate}
    \item \textbf{Preemptive response:} Predicting the end of the turn and starting to reply before it is completed. For example, participants preemptively gave answers to the interlocutor’s questions as in ``A: Do you remember who the movie direc--'' ``B: You mean Bong Jun Ho?''
    \item \textbf{Backchanneling:} Showing one is paying attention or giving instant agreement on others’ perspective. For example, participants gave backchanneling to the interlocutors as in ``A: Today I went to--'' ``B: yeah.'' 
\end{enumerate}

Second, we observed that participants frequently engaged in \textbf{deletion} behaviors. Specifically, to resolve interruption from their interlocutor, participants deleted their typed messages. This happened when participants encountered simultaneous typing by interlocutors. As mentioned in Section \ref{related-work-overlap} about competitive overlap, the concept of interruptions necessitates some level of interpretation, where one participant is perceived as violating the other's right to speak. We interpreted this deletion behavior as the resolution mechanism for interruption, to determine which speaker should continue. All participants demonstrated and acknowledged this counteracting behavior to interruptions. Participants reported their reasons as follows, which are related to competitive overlap (Section \ref{related-work-overlap}).

\begin{figure*}[t!]
    \centering
    \subfloat[UI showing typing]{\includegraphics[width=0.48\linewidth]{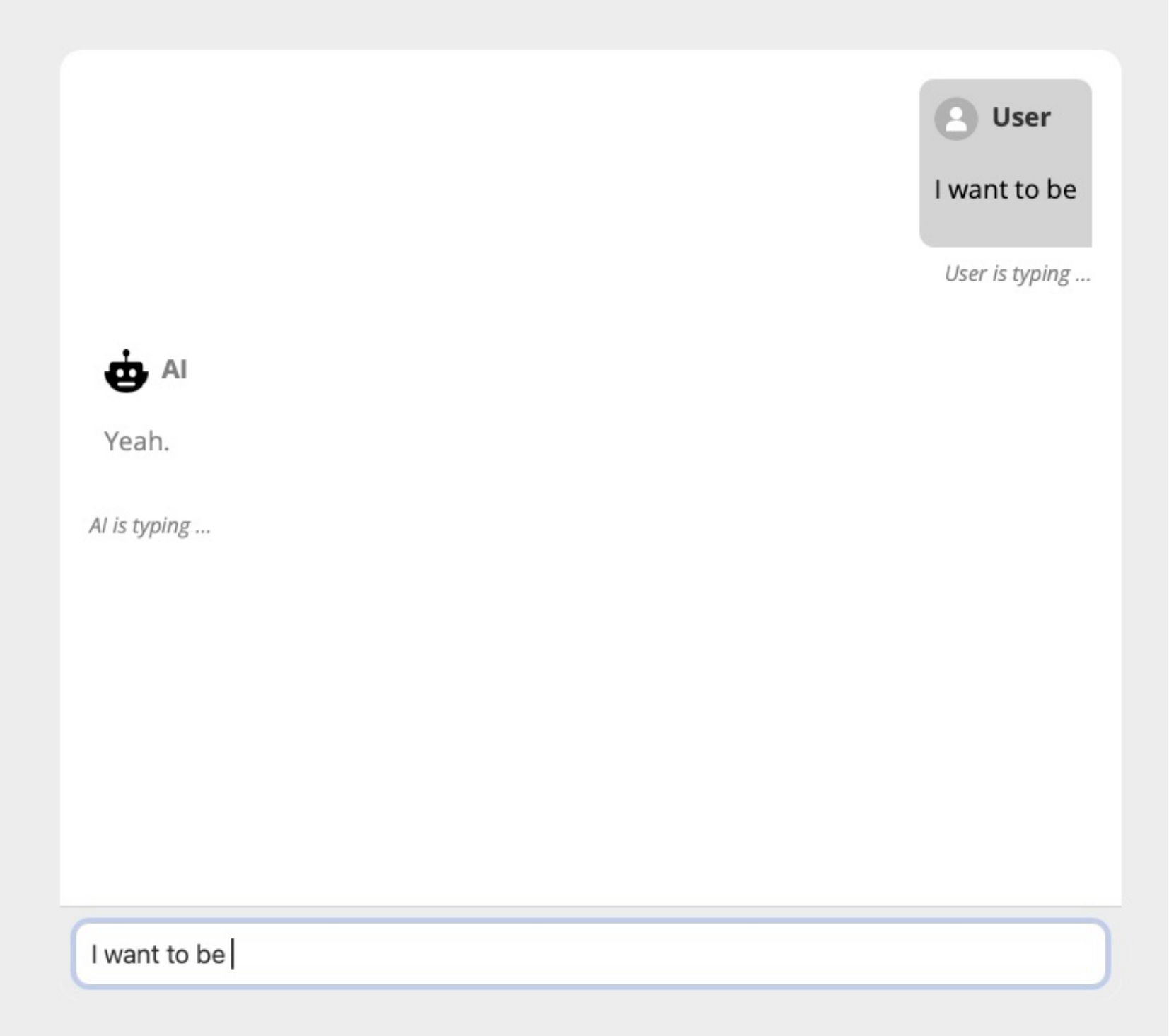}}\hspace{-0.5em}
    \subfloat[UI showing sent]{\includegraphics[width=0.48\linewidth]{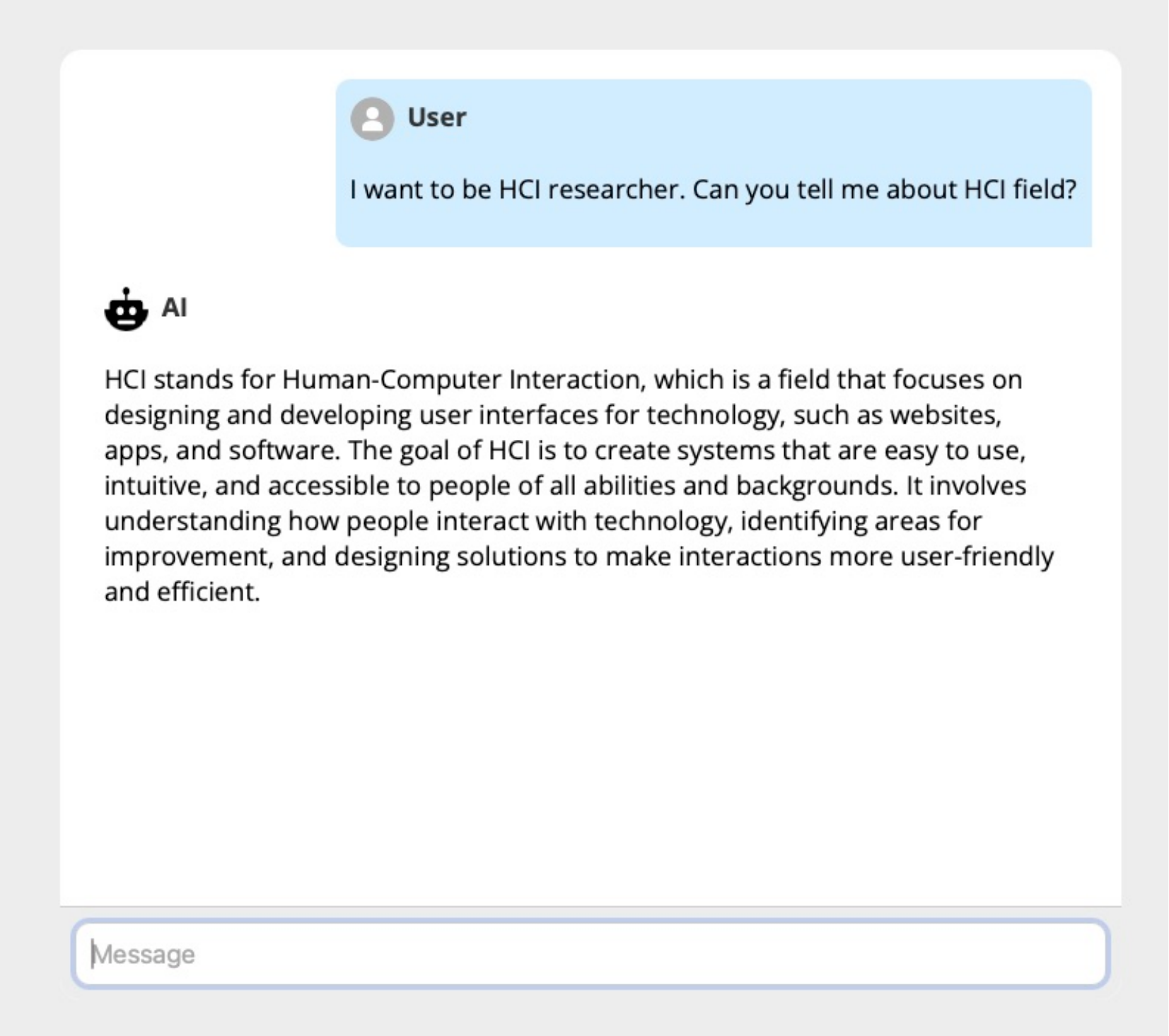}}
    
    \caption{Visually changed UI from typing status to sent status.}
    \label{fig:typing-in-webinterface}
\end{figure*}

\begin{enumerate} 
    \item Adjusting responses based on the interlocutor’s actions, such as transitioning topics when there is a mismatch or addressing questions and refutations during simultaneous typing. 
    \item Removing brief real-time feedback, including backchanneling cues, typos, or profanity. 
\end{enumerate}

In addition, participants perceived conversations using the probe as authentically similar to a real conversation. They noted that the flow of conversation with overlapping was uninterrupted, enhancing the presence of the interlocutor and fostering greater engagement. \textit{``It made me focus more on the chat because I could see what the other person was typing (and they might even delete it).'' (P3)}; \textit{``When the content I was about to type matched what the other person was typing, it felt like a boost in closeness.'' (P5)} The prevailing sentiment was that overlapping effectively promoted the exchange of opinions: \textit{``It felt like the limitations of online discussion were reduced.'' (P1)} 

However, certain participants experienced a psychological burden due to the transparency of their thought processes while typing \cite{podlubny2017synchronous}. \textit{``Since everything I typed was visible to the other person in real time, even what I typed unconsciously, I became more cautious.'' (P8)}; \textit{``If I had to chat for an extended period with this interface, I think I would feel fatigued, as if my initial thoughts were being monitored.'' (P4)} Some participants expressed a preference for using this interface exclusively in intimate relationships: \textit{``I would use it with close friends, but probably not with people I am not as familiar with.'' (P12)}  

In conclusion, these findings reveal that people instinctively engage in overlapping during text-based interactions -- something traditional chatbot systems don’t allow. We have grown so accustomed to strict turn-taking with chatbots that we may not have realized what has been missing. When given the chance to overlap, people naturally embrace it, opening up possibilities in chatbot design. This naturally occurred conversational behavior presented a new technical challenge where text-based chatbot cannot naturally overlap people, which we solved by finetuning LLM with publicly available datasets customized for overlapping. 

\subsubsection{Comparison to Previous Literature}
This section contextualizes our findings in relation to existing research on 1) real-time human-human text interactions and 2) speech-based dialogues. 

First, while numerous study have studied real-time messaging, the most comparable study \cite{kim2017applying} identified three key findings: a) reduced silence, b) real-time text serving as a nonverbal expression channel contributing to a sense of remote presence, and c) new texting techniques such as intentional delay, instant backchanneling, and deletion to convey irony (e.g., ``I don’t (delete) really like (rewrite) that idea''). While our findings align with this study, our focus was on identifying actionable behaviors that can be performed by an LLM. In addition to backchanneling, which is supported by previous literature on human-AI interactions \cite{DBLP:conf/chi/DingKHWFMM22, DBLP:conf/hri/ParkGLB17}, we identified textual preemptive responses as a key behavior, highlighting their potential role in conversations between a questioning human and a responding LLM. Furthermore, techniques like ironic deletion or intentional delay may not be suitable for an LLM in these contexts.

Second, we observed both similarities and differences compared to prior studies on overlapping in human-human speech-based dialogue. Participants exhibited cooperative overlaps, such as preemptive responses and backchanneling, and resolved interruptions as competitive overlaps, consistent with prior studies. Unlike speech-based interactions, where semi-verbal cues like pitch and speed enhance expression, text-based overlap-capable interactions lack these features but introduce a unique channel through message deletion. \textit{``I made a joke about an anime song and then deleted it. It was interesting because it is the kind of joke that would not be as easy to make without this kind of interface.'' (P3)} This behavior reflects distinct mental models for deciding what to keep or remove from conversation logs and highlights differences in how interruptions are resolved in text versus speech, suggesting opportunities for future research.

\section{AI Chat Prototype Development}\label{3-0-system-building}
The results of our formative study showed that people naturally engaged in overlapping and deletion. To replicate these behaviors in an LLM, we developed OverlapBot (RQ2). OverlapBot is an AI chat web prototype where the chatbot uses backchanneling and preemptive answering to overlap with the user's typing. 

We developed the OverlapBot by finetuning an open-source LLM with customized datasets, enabling both the chatbot and users to engage in overlapping conversations. Unlike previous real-time text interfaces on human-human interactions, our AI-driven interface emphasizes human-AI interaction with overlapping capabilities. The web server for OverlapBot was built using Python Flask to manage interactions and data handling efficiently. We selected the Llama3-8B model as the base for finetuning our chatbot. To create a customized dataset, we combined the conversational SwitchBoard dataset with an instruct-tuning dataset. For further details on the dataset preparation, please refer to Appendix \ref{Appendix-datamanipulation}.

Three key design considerations were essential to facilitate this overlapping, which we will introduce in the following sections.

\subsection{Real-time Typing}

Similar to the research probe used in the first formative study, this interface displays participants' typing activities in real-time, allowing others to see their input as it happens.

Displaying participants' typing activities is a foundational element of OverlapBot, as it enables the sharing of context in real-time, closely mirroring the interactions of spoken conversation. By making participants share the context in real-time, the interface creates opportunities for overlap. For example, as illustrated in Figure \ref{fig:typing-in-webinterface} (a), if a user types ``I want to be'' and the chatbot writes overlapping message by saying ``Yeah,'' the overlap occurs at the word ``be'' in the user's utterance. Without real-time typing display, identifying such overlapping points would not be possible.

As shown in Figure \ref{fig:typing-in-webinterface}, we implemented a visual change from typing status to sent status. User's message bubble changes color from gray to light blue to indicate that they have finished typing. Similarly, the chatbot's status changes from typing to sent, with the text color shifting from gray to black. When participants are typing, texts of \textit{`... is typing'} are displayed below the messages. We opted not to place the chatbot's text in a message bubble, as a dynamically changing message box could become visually tiring if the chatbot's response, shown during typing, becomes too lengthy. In addition, we did not include a visible `Send' button, opting instead to implement message submission via pressing the Enter key. This decision was informed by our observation that participants in the formative study tended to press enter to send their messages rather than moving their mouse cursor. 

We implemented real-time typing by systematically collecting all of the user's keystrokes and displaying them on the web interface, while simultaneously showing the chatbot's response one character at a time.

\subsection{Overlap}

\begin{figure*}[t!]
    \centering
    \subfloat[Backchanneling]{\includegraphics[width=0.48\linewidth]{figure-webinterface-interactionflow1.pdf}}\hspace{-0.5em}
    \subfloat[Preemptive Answering]{\includegraphics[width=0.48\linewidth]{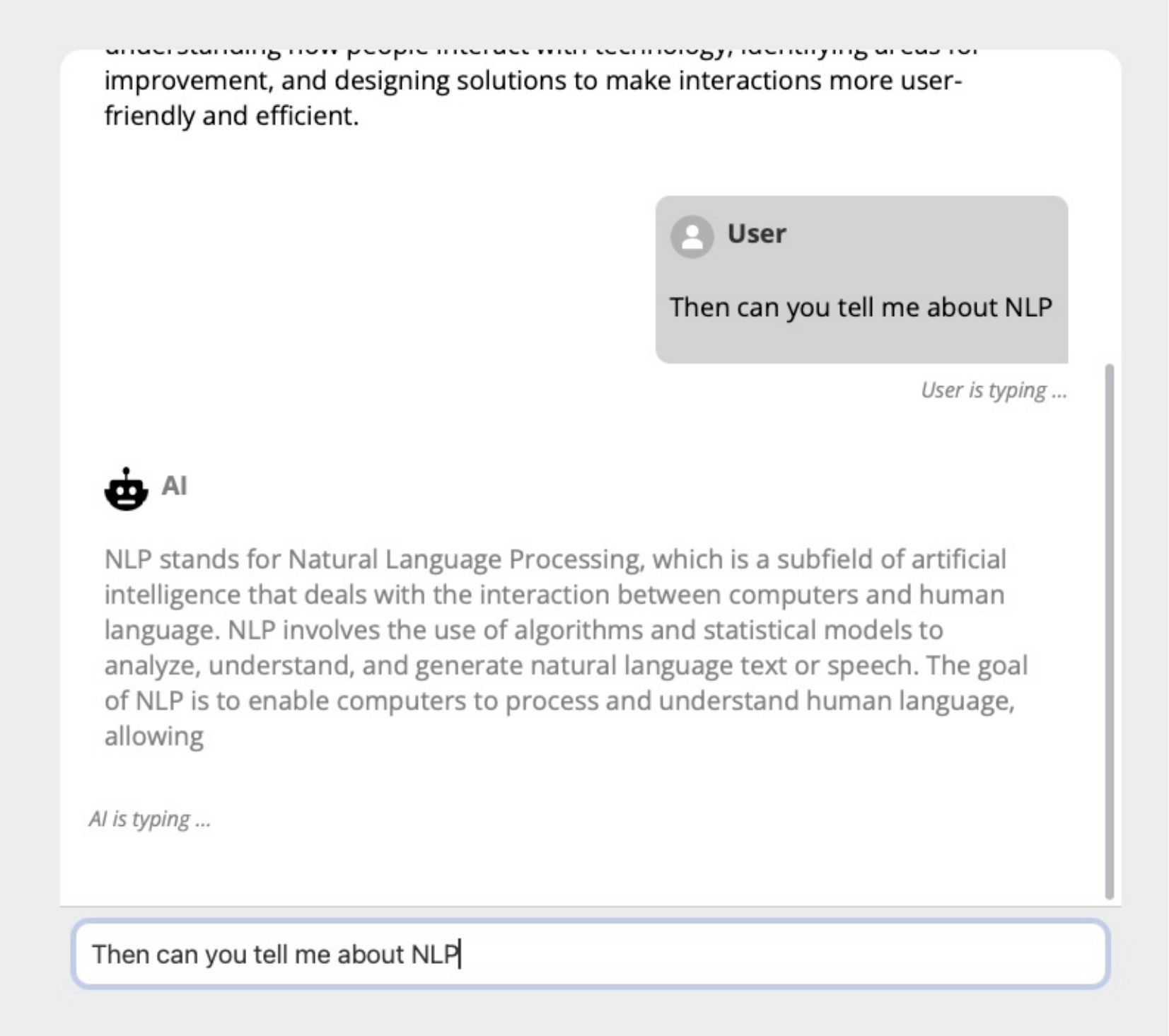}}
    
    \caption{
    Examples of the types of overlap by OverlapBot. While the user is typing, OverlapBot can provide listener cues indicating attention (Backchanneling) or generate a response before the user finishes typing (Preemptive Answering).
    }
    \label{fig:overlapping-in-webinterface}
\end{figure*}

The real-time visibility of typing activities allows both users and the chatbot to initiate overlapping responses before the other party presses `send'. As shown in Figure \ref{fig:overlapping-in-webinterface}, the chatbot can produce two types of overlap: 1) preemptive answering and 2) backchanneling. These behaviors are informed by findings from the formative study.

With preemptive answering, the chatbot can generate responses based on the user’s input before they complete it or press Enter. Through backchanneling, the chatbot offers immediate listening cues, allowing it to acknowledge the user without disrupting the conversation's flow.

Unlike other design considerations that could be systematically implemented, overlapping functionality depends on the chatbot's ability to create overlaps. To achieve this, we trained an open-source LLM with custom-built datasets. Our model outperformed baseline models, including GPTs. Details on how we equipped the chatbot with overlapping capabilities are provided in Appendix \ref{sec:model-building}.
 
\subsection{Interruptions and deletions}
\begin{figure*}[t!]
    \centering
    \subfloat[Before Interruptions]{\includegraphics[width=0.33\linewidth]{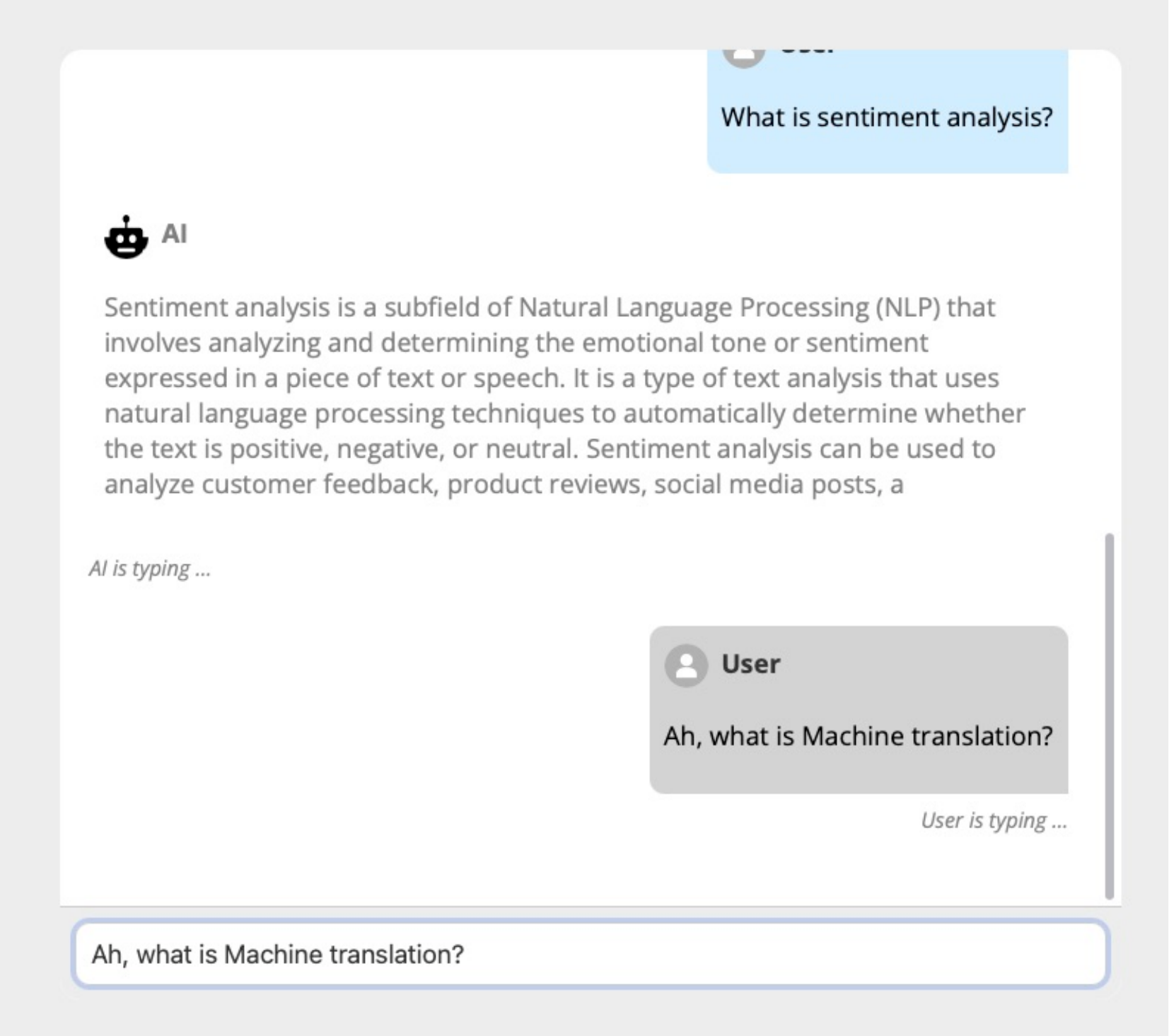}}\hspace{-0.5em}
    \subfloat[Interrupted]{\includegraphics[width=0.33\linewidth]{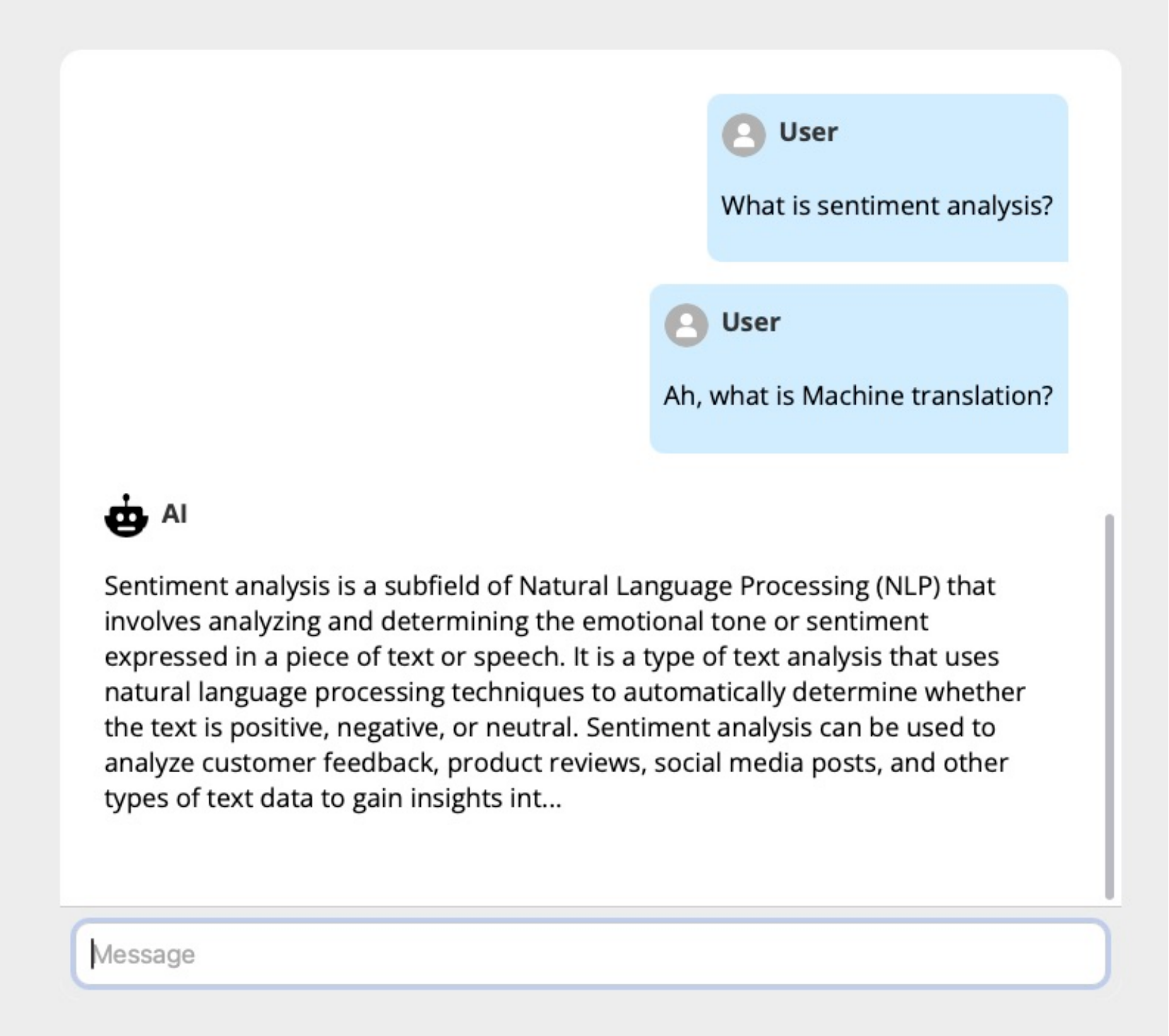}}
    \subfloat[After Interruptions]{\includegraphics[width=0.33\linewidth]{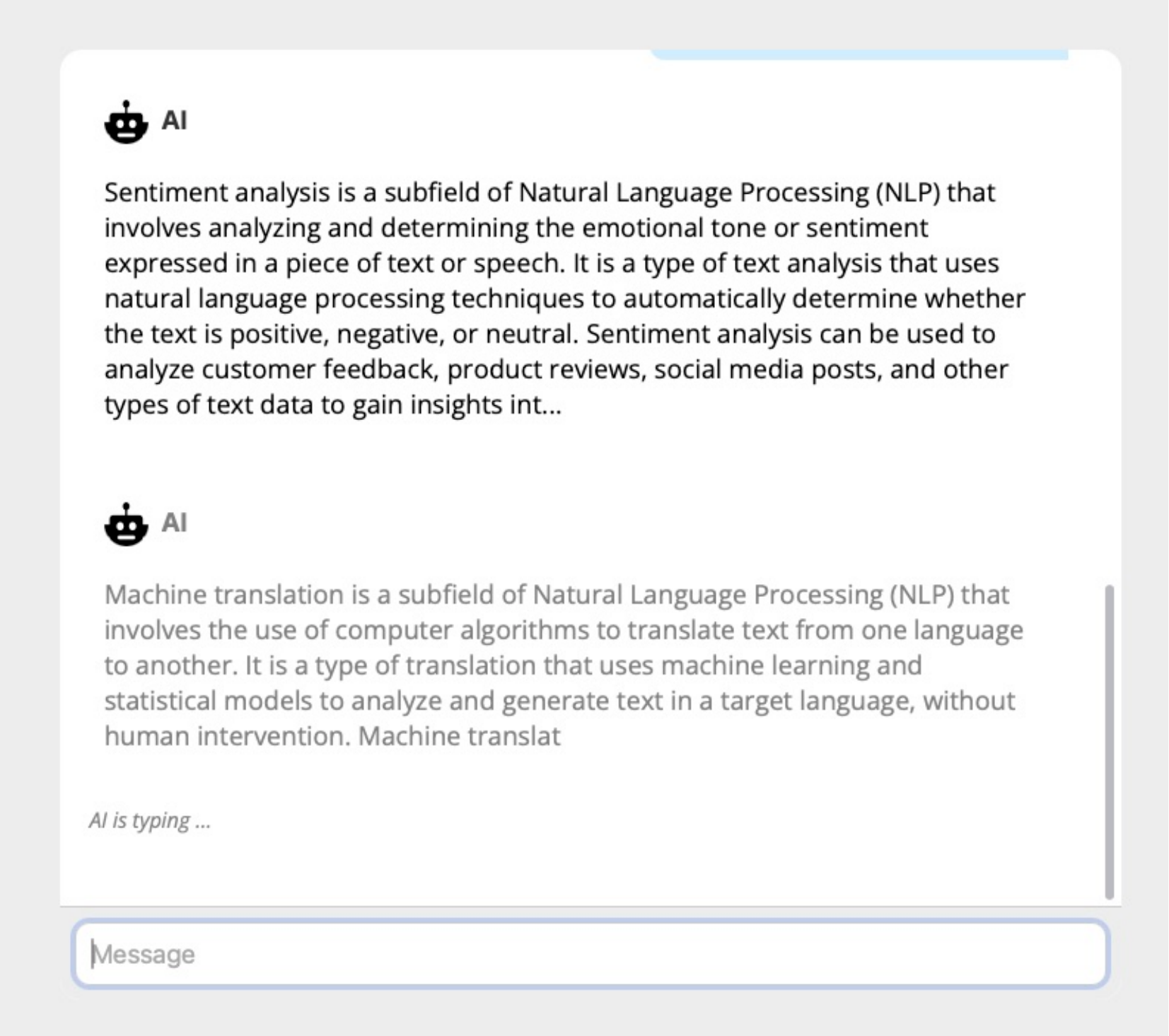}}
    
    \caption{The chatbot regenerates an answer based on the interruption, presented in a timely order from (a) to (c): Before the interruption (a), the chatbot is generating an answer about sentiment analysis. The user interrupts by asking about machine translation (b). After the interruption (c), the chatbot generates a new answer about machine translation.}
    \label{fig:deletion}
\end{figure*}

Motivated by participants' deletion behavior as a way to handle interruptions, the chatbot is programmed to delete its prior response, including backchanneling, when interrupted by the user, and generate a new one. However, constantly deleting and regenerating entire messages after each interruption could lead to user fatigue. To address this, we implemented a solution where, if the chatbot's response exceeds 130 characters, it adds `...' at the end of the interrupted message, signaling that the response is still in progress (Figure \ref{fig:deletion}). Through empirical testing, we found that 130 characters maintain the optimal conversation flow. The order of message boxes is determined by the sequence of `Send’ actions. If both the user and the chatbot are typing simultaneously and the user clicks `Send’ first while the chatbot continues typing, the user's message box is placed above.

In our work, we categorized user overlaps with the chatbot's typing as interruptions, based on the inherent power dynamics and interaction patterns typically seen between humans and AI \cite{DBLP:conf/chi/Sarkar23}. In these interactions, the AI is generally positioned as a responsive entity, focusing on topics initiated by the human user. Given this context, when users overlap with the AI's responses, it often signifies an attempt to regain control of the conversation rather than a collaborative effort, making these overlaps more similar to interruptions than cooperative exchanges.  

\section{Study 2 - Understanding Human Perceptions on LLM Overlap and New Textual Interactions}\label{user-study}
To explore how people perceive overlapping interactions between humans and an LLM, and to identify new interaction patterns (RQ3), we conducted an experiment using OverlapBot. The goal of this study was to uncover new design opportunities for enabling overlap in text-based interactions with LLM. 

For comparison, we implemented a conventional chat system where neither the users nor the chatbot could see each other's typing. In this system, we employed the basic, unfinetuned Llama3-8B model. 

\begin{table}[t!]
    \centering
    \caption{Quantitative comparison of conventional chatbot and OverlapBot in our study. Overlap Ratio represents the percentage of total conversation time where simultaneous keystrokes occurred between the User and OverlapBot.}
    \label{tab:quantitative}
    \renewcommand{\arraystretch}{1.1}
    \resizebox{\columnwidth}{!}{%
    \begin{tabular}{lcrr}
    \toprule
    \textbf{Metric}               & \textbf{Role}       & \textbf{Conventional}         & \textbf{OverlapBot}          \\ \midrule
    \multirow{2}{*}{\textbf{Message Length}} 
                                  & User                & 62.36 {\scriptsize (±22.49)}  & 43.18 {\scriptsize (±12.74)} \\
                                  & Chatbot             & 177.64 {\scriptsize (±34.65)} & 133.40 {\scriptsize (±42.19)} \\ \midrule
    \multirow{2}{*}{\textbf{Total Turns}} 
                                  & User                & 7.56 {\scriptsize (±2.59)}    & 13.00 {\scriptsize (±3.93)}  \\
                                  & Chatbot             & 7.33 {\scriptsize (±2.40)}    & 16.89 {\scriptsize (±7.19)}  \\ \midrule
    \multirow{2}{*}{\textbf{Turns per Minute}} 
                                  & User                & 1.28 {\scriptsize (±0.45)}    & 1.93 {\scriptsize (±0.82)}   \\
                                  & Chatbot             & 1.25 {\scriptsize (±0.45)}    & 2.48 {\scriptsize (±1.33)}   \\ \midrule
    \textbf{Overlap Ratio}        &                    & -                             & 6.0\% {\scriptsize (±3.0\%)} \\ \midrule
    \multirow{2}{*}{\textbf{Deletes per Minute}} 
                                  & User                & -                             & 11.10 {\scriptsize (±6.62)}  \\
                                  & Chatbot             & -                             & 2.98 {\scriptsize (±1.70)}   \\ \bottomrule
    \end{tabular}%
    }
\end{table}

A total of 18 participants were recruited by voluntarily responding to the experiment participation post on the university’s website. 10 of them were South Koreans, 6 of them Indonesians, 1 of them Nepali, and 1 of them is Vietnamese. Their average age was 23 ($SD$=2.42), and 9 of them were female and 9 were male. They all self-reported frequent usage of the OpenAI chatGPT website. As compensation for their participation, all participants were paid 50K KRW. Each experiment lasted approximately 60 min on average. All sessions were conducted remotely using Google Meets with audio and video recordings and were conducted in Korean or English, based on the nationality. The overall procedure of our study was conducted after obtaining IRB approval from the university.

Before the experiments began, participants received a detailed explanation of how to use OverlapBot and the conventional chat system. The tutorial introduced key functionalities of OverlapBot, such as its ability to display real-time typing and provide understanding reactions (e.g., “yeah”) or answers before the participant’s utterance was complete. Participants were also instructed on how to interrupt the chatbot’s response. For the conventional chat system, they were informed that neither they nor the chatbot could see each other’s typing in real-time. During the tutorial, participants were given examples of potential conversation topics, such as discussing hypothetical scenarios like “Would you rather speak every language or communicate with animals?” or “Would you rather die in 20 years with no regrets or live to 100 with a lot of regrets?” The explanation and tutorial session was conducted for approximately 10 minutes. 

Following the tutorial, each participant engaged in a 10-minute conversation with the conventional chat system and then with the OverlapBot, with the order randomized. Participants were free to choose any topic including the hypothetical scenarios for their interactions in English, ensuring that the conversations were natural and varied. After completing the interactions, participants were asked to fill out an open-ended survey and participate in a semi-structured interview to gather qualitative feedback on their experience. The survey and interview included questions designed to explore participants’ perceptions and preferences regarding the two chatbots. Key questions addressed the main differences participants noticed between the OverlapBot and conventional chatbots, their overall impressions of each chatbot, and specific aspects of the OverlapBot that they found most useful or convenient. Participants were also asked to indicate which interface they preferred and to explain their reasons. Additionally, the survey inquired about any difficulties or discomfort experienced while using the Overlapbot.

We collected four types of data: open-ended survey responses, interview responses, recorded videos, and conversation logs. We analyzed participants' conversation logs to conduct a quantitative comparison between OverlapBot and a conventional chatbot. We utilized open-ended survey responses and transcribed interview responses to analyze general impressions of OverlapBot compared to the conventional chat system. For the analysis, thematic analysis \cite{boyatzis1998transforming} was conducted by three authors. We repeatedly observed recorded videos to learn new interaction patterns users showed using OverlapBot. Three authors conducted a thematic analysis \cite{boyatzis1998transforming} of the transcribed interview and open-ended survey responses to gather insights on participants' impressions of OverlapBot. \label{measures}

\subsection{Findings}
Table \ref{tab:quantitative} presents the quantitative results, highlighting that OverlapBot, in comparison to the conventional chatbot, facilitated shorter message lengths and a higher number of turns exchanged between the user and the chatbot. Here, turns are calculated based on Send actions, not typing status. Notably, the chatbot in the OverlapBot interface sent messages more frequently than the conventional chatbot, indicating its ability to provide more information within the same timeframe. Interestingly, the ratio of turns exchanged between the user and the chatbot, which was nearly a balanced exchange of turns in the conventional interface, shifted in the OverlapBot interaction. This shift could be attributed to OverlapBot's backchanneling behavior, which might not have elicited responses from users. Additionally, users deleted messages more frequently than OverlapBot, possibly due to revising their written content before resending it to the LLM, or intentionally removing their input to avoid leaving their words in the conversation logs.

Supplementary qualitative analyses are presented in the following sections.

\subsubsection{General Impressions}

\renewcommand{\arraystretch}{1.5} % 행 간격 조정
\setlength{\tabcolsep}{4pt} % 열 간격 축소
\begin{table*}[t!]
\caption{Perception on OverlapBot}
\label{tab:themes}
\centering % 테이블을 중앙 정렬
\begin{tabular}{>{\centering\arraybackslash}m{3cm}|>{\centering\arraybackslash}m{3cm}|>{\arraybackslash}m{9cm}} % 열 너비 명시적 설정
\toprule
\textbf{Main theme} & \textbf{Sub theme} & \textbf{Quotes} \\
\midrule
\multirow{2}{*}{\textbf{Human-like}} 
    & Communicative & \textit{``Unlike conventional chatbot, it felt more like having a conversation with someone who has emotions. Instead of just outputting data, it gave a stronger impression of actually conversing, so I would recommend it to someone who needs a conversational partner.'' (P16)} \\
    & Immersive & \textit{``How I feel engaged into conversation was much better in OverlapBot, since humans used to talk in real-time for a very long time.'' (P8)} \\
\midrule
\textbf{Speedy Interaction} & Efficient & \textit{``I liked the fact that the conversation was fast-paced. I felt that the pace of the conversation was relatively quick because it started preparing responses as soon as I said something. As someone who generally prefers things to move quickly, this was a very convenient aspect.'' (P16)} \\
\midrule
\textbf{Brief Response} & Less structured & \textit{``The conventional chatbot gives more well-constructed, elaborate answers, while OverlapBot is significantly more impulsive, and the responses are catered to shorter, general responses.'' (P18)} \\
\bottomrule
\end{tabular}
\end{table*}
 
In this experiment, we found that participants had three general impressions of OverlapBot compared to the conventional chatbot (Table \ref{tab:themes}). 

\paragraph{Human-like Interaction} The most common impression of OverlapBot was that interactions with the chatbot felt similar to conversing with a real person. Participants specifically noted that the chatbot felt more communicative and immersive compared to the conventional chatbot: \textit{``The engagement between the user and the chatbot was significantly more immersive, giving the feeling of actually talking to a real person.'' (P8)}; \textit{``The immediate response that the chatbot provided made the conversation feel more like a natural exchange.'' (P2)} It is notable that participants described OverlapBot as like talking to a friend and the conventional chatbot as like using a search engine. These comments reflect how OverlapBot fosters a friendly and engaging conversational environment, akin to talking with a friend.

\paragraph{Speedy Interaction} The next common impression on OverlapBot was speedy interaction. As the chatbot can give preemptive responses during users' typing and users can interrupt the chatbot, the conversation itself can be fast-paced. \textit{``Being able to get the response to the question before completely asking it was time efficient.'' (P10)} Real-time messaging also contributed to the speedy interaction. \textit{``With the conventional chatbot, I have to wait while it completely generates the message before sending it. With OverlapBot, I could see what it's typing as it writes, allowing me to read it simultaneously, which felt more time-efficient.'' (P12)} However, some participants experienced delays in getting responses due to technical issues. This is caused by the technical limitation which we tackle in Section \ref{limitations}. Despite the limitation, the overall perception was that OverlapBot allowed for quicker and more immersive communication than with the conventional chatbot.
    
\paragraph{Brief Response} While the increased speed was generally seen as a positive, some participants observed that the OverlapBot’s responses were shorter and less structured. Specifically, the conventional chatbot often gave detailed answers, while OverlapBot gave concise ones. \textit{``It was quite unsuitable for more elaborate discussions, the responses were very brief.'' (P18)}; \textit{``While the responses (of OverlapBot) were faster, they sometimes felt a bit abrupt or lacking in detail.'' (P7)} This brevity was likely influenced by a conversation dataset that OverlapBot was trained on. In the conversation dataset, many utterances are relatively brief, showing that this is not a problem of overlapping itself.

\subsubsection{New Textual Interactions with LLM}

\begin{figure*}[t!]
    \centering
    \includegraphics[width=1\textwidth]{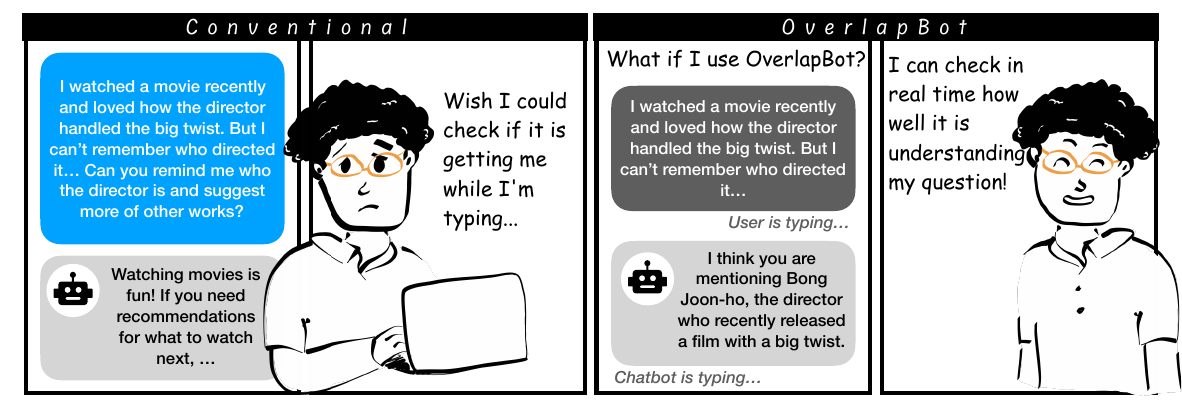}
    \caption{First new human-LLM interaction: The user checks OverlapBot's active listening through its preemptive answering.}
    \label{fig:overlapanswering}
\end{figure*} 

In this experiment, we observed three new types of interactions emerging in human-LLM interactions, including users' behavioral and attitudinal responses to the LLM. Although we designed the chatbot with actionable behaviors - preemptive answering, backchanneling, and deleting -, we did not fully anticipate how participants would embrace and respond to these behaviors. Responding to the LLM's actionable behaviors, participants showed behaviors and attitudes that closely mirrored those observed in human-human conversational interactions. This distinguishes themselves from the interactions typically seen with traditional turn-taking chatbots.

\paragraph{Users interpret an LLM’s preemptive responses as a form of listening. } Participants naturally responded to OverlapBot's preemptive answering as if it were a form of active listening (Figure 6), a behavior that resembles human-human conversation patterns. In human conversations, people signal their attention to the conversation through active listening cues \cite{heinz2003backchannel}. Similarly, participants used OverlapBot's preemptive answering to assess how well it understood the topic and how actively it was engaging in the conversation. This behavior differed from interactions with the conventional chatbot, where participants waited for the chatbot to respond to their questions rather than actively checking its engagement. \textit{``With the conventional chatbot, I have to complete a full sentence to get a response. However, with OverlapBot, it responded as soon as it understood enough of the sentence. This made the interaction more convenient, as I could decide whether to finish typing, edit, or stop based on the chatbot's immediate response.'' (P5)}; \textit{``Even if I'm still thinking about my question and have not finished typing, the chatbot already provides feedback. This mimics the flow of a conversation between humans, where responses often come before a statement is fully formed.'' (P11)} It was often seen as a way to speed up the conversation by providing immediate feedback. However, some participants found frequent or poorly timed overlaps to be intrusive, disrupting their train of thought: \textit{``It feels like talking to an annoying friend who interrupts and starts saying what they want to say before I've finished speaking.'' (P13)} 
    
\paragraph{Users recognize an LLM’s backchanneling but without verbal acknowledgment. } 

\begin{figure*}[t!]
    \centering
    \includegraphics[width=1\textwidth]{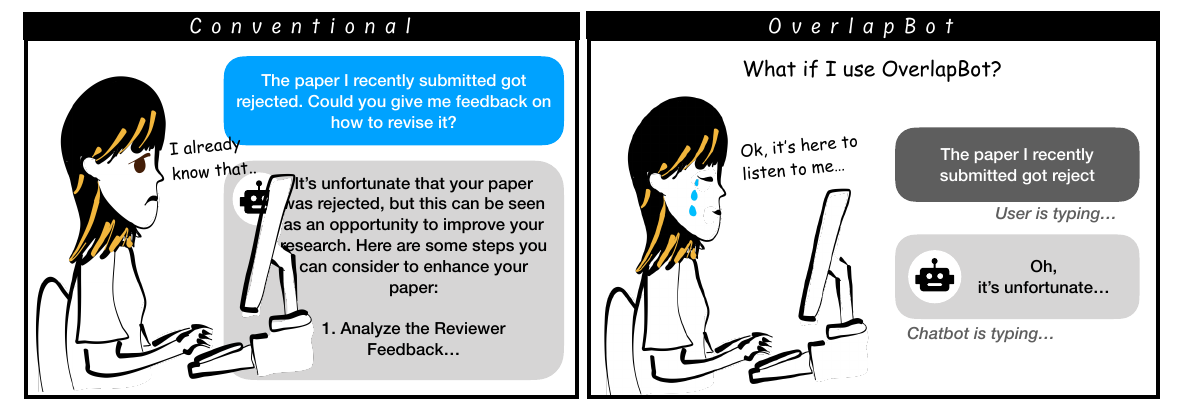}
    \caption{Second new human-LLM interaction: The user makes no verbal reaction to OverlapBot's backchanneling but acknowledges its presence.}
    \label{fig:backchanneling}
\end{figure*}  

Although participants did not provide a verbal response to OverlapBot's backchanneling cues, they still recognized its presence, perceiving it as a sign of attentiveness (Figure \ref{fig:backchanneling}). This mirrors human-human interactions, where backchanneling often occurs without explicit acknowledgment but still contributes to a sense of engagement. In human-human conversations, backchanneling does not require a direct response, as it is not typically considered ``turns'' in conversation (Section \ref{related-work-overlap}). Similarly, participants did not verbally acknowledge the chatbot's backchanneling with comments like, ``Oh, you are reacting to me!'' Yet, they still perceived the chatbot's presence in the chat, much like in human-human interactions. This behavior differed from interactions with the conventional chatbot, where participants would receive backchanneling only as part of the chatbot's full response: \textit{``Both chatbots add backchanneling and then provides its intended response, but the difference is whether that response comes in one complete sentence, like with a traditional chatbot, or bit by bit in real-time, like with OverlapBot. In that sense, I feel that the OverlapBot is better at giving the feeling of genuine communication.'' (P8)} Many participants noted that OverlapBot's backchanneling made the chatbot seem more attentive and responsive, enhancing its overall presence in the conversation. \textit{``The chatbot's immediate responses gave the impression that it was actively listening, which made the conversation feel more engaging'' (P8)}; \textit{``Seeing real-time responses like `yeah' made me feel like I was not really talking to an AI.'' (P2)} However, for participants primarily focused on using the chatbot for information retrieval, these reactions were less appreciated, as they did not directly contribute to the information exchange they were seeking: \textit{``I don't really like my chatbot reacting when I am asking a question.'' (P6)}; \textit{``Real-time reactions did not really feel like a significant advantage to me.'' (P4)} 

\begin{figure*}
    \centering
    \includegraphics[width=1\textwidth]{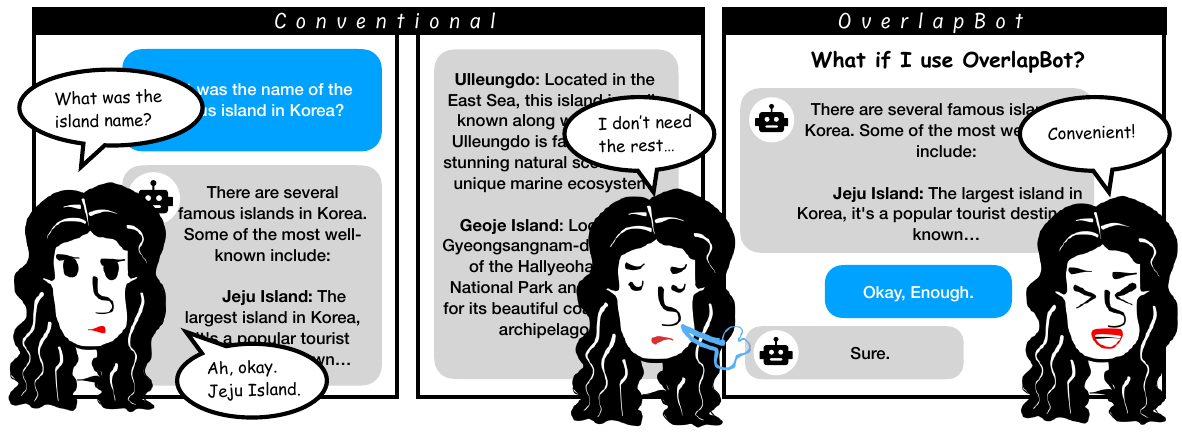}
    \caption{Third new human-LLM interaction: The user makes short interruption commands to OverlapBot.}
    \label{fig:interruptions}
\end{figure*}

\paragraph{Users interrupt an LLM with brief commands.} Participants used short, direct commands to interrupt OverlapBot, prompting it to stop typing (Figure \ref{fig:interruptions}). This pattern mirrors human-human interactions, where brief interruptions help manage the flow of conversation. In human-human conversations, people take turns by interrupting their interlocutor's speech, causing the other person to stop talking \cite{skantze2021turn}. Similarly, when participants wanted to interrupt the chatbot, they often did so abruptly with short commands like ``stop'' or ``okay,'' before crafting their next prompt. This behavior differed from interactions with the conventional chatbot, where participants prepared a new prompt in detail while the chatbot was still generating a response.  \textit{``Stopping OverlapBot's chatbot and asking a completely different question felt very easy and convenient.'' (P6)}; \textit{``It is comfortable that I can make it stop responding when I want and switch to another topic.'' (P9)}; \textit{``OverlapBot felt like I was actually having a conversation with someone. This was because, just like when talking to a real friend, I could predict what it was going to say after it deleted its response when I interrupted its typing.'' (P2)} This pattern indicates that when an interruption point arises, participants are more likely to disregard the ongoing response and initiate their own input. 

\section{Design Insights and Discussion} 
\label{discussion}
Based on our findings and insights from the two studies, we propose the following design insights with discussion for handling overlapping in text-based interactions (RQ4). The goal of this section is to introduce new opportunities for developing overlap-capable chat systems to researchers and designers \cite{DBLP:conf/chi/AmershiWVFNCSIB19, DBLP:conf/chi/WeiszHMHMG24}. 

\subsection{Overlap for Interactive and Multiparty Situations} 
Overlap may be less effective in non-interactive or single-turn scenarios where user engagement does not directly influence the LLM’s output. For instance, when a user merely reads a model’s response without providing input or modifying its generation, overlap could disrupt focus or clarity instead of enhancing productivity. Examples of such scenarios include single-turn tasks like math calculations or closed question answering, where the primary goals are accuracy and single-response clarity rather than maintaining the fluidity of dialogue.

Conversely, we believe overlapping features are well-suited for interactive and iterative conversational situations, where they can boost both engagement and productivity. These scenarios involve users actively interacting with the LLM, not merely as evaluators but as participants in iterative processes involving repeated questions or multi-turn exchanges \cite{DBLP:journals/tmlr/0002SHTDPGLLRWK23}. Overlapping interactions can also be effective in multiparty settings. Our initial research has identified patterns like backchanneling and terminal overlap, but the potential to model other forms of overlap — such as choral talk, where multiple participants speak simultaneously — introduces further opportunities for multiparty chatbot design. In these complex scenarios, where coordinating multiple speakers is particularly challenging, a chatbot capable of accurately determining when to interrupt or contribute could facilitate smoother interactions. 

\subsection{Overlap Across Tasks and Relationships}
Interactive situations can span a wide range of tasks, from goal-oriented tasks (e.g., question answering, text summarization, crossword puzzles) to open-ended tasks (e.g., social dialogue, metaphor generation) \cite{DBLP:journals/tmlr/0002SHTDPGLLRWK23}. However, how overlapping interactions unfold in detail for each task remain open questions for future research. For instance, while social dialogue may naturally support overlapping, tasks like text summarization with longer user inputs might be less suitable. Some users in our study highlighted that overlapping could be effective in brainstorming tasks. Additionally, in time-sensitive scenarios like disaster management, participants recognized the potential of OverlapBot’s real-time responsive assistance to provide critical support. These findings suggest that task-specific contexts significantly influence the effectiveness and appropriateness of overlapping features, which we leave to the future work. 

In addition to task types, the conversational relationship between humans and AI also requires further exploration. Participants in our formative study observed that the transparency of typing might feel more appropriate in casual relationships, such as with close friends, but less suitable in hierarchical or unfamiliar relationships: \textit{``I would use it with close friends, but probably not with people I am not as familiar with.'' (P12)} This suggests that the relational context of human-AI interactions — whether focused on companionship, practical assistance, or other roles — may influence how overlapping features are perceived and received. For instance, socially isolated individuals, such as the elderly or those living alone, may appreciate OverlapBot’s overlapping features as part of its role as a conversational partner. On the other hand, users engaging with AI in professional or hierarchical settings may favor stricter turn-taking norms. These nuanced preferences highlight the need to design overlapping interactions that are sensitive to the role and context of the relationship.

\subsection{Rethinking the Necessity of Prompting Design}
Numerous studies have shown that LLMs produce varying outputs based on the prompts they receive, prompting users to carefully craft precise prompts. Our findings suggest that overlap may reduce the need for highly detailed prompts. By observing the user’s input in real time as they type, the LLM can infer intent without relying on a fully developed prompt. As the LLM anticipates the user’s intended response, users can provide immediate confirmation or correction. However, while some participants appreciated this as a convenient and effective feature, others found it uncomfortable, viewing the typing process as a critical step for clarifying and organizing their thoughts. This feedback indicates that overlapping interface should offer users control, enabling them to adjust the visibility of their typing to match their interaction preferences.

\subsection{User-Customizable Overlap}
When designing overlapping chatbots, it is essential to consider user preferences and provide adjustable settings that accommodate diverse interaction styles. Some users, particularly those accustomed to signaling the end of their turn with the Enter key, may find the chatbot’s proactive behavior intrusive or disruptive. To address this, the chatbot must carefully determine the right moment to offer a preemptive response, ensuring users feel they have communicated enough before being interrupted. As one participant shared: \textit{``I wish it would let me finish what I have to say. (...) I feel like I have to finish speaking quickly or say something just to keep up, and that made me feel uncomfortable and uneasy.'' (P16)}

The absence of non-verbal cues in text-based interactions complicates this further. As another participant noted: \textit{``In human conversations, you can usually guess from my facial expressions or tone, but here, it only relies on the text, so I thought there might be more room for error.'' (P12)} One possible solution is to adjust overlap frequency based on the user’s typing speed. For example, slower typists may benefit from more frequent overlaps to maintain flow, whereas faster typists might find them disruptive. 

\subsection{Culturally Adaptive Overlap}
When designing overlapping chatbots, it is essential to account for cultural differences, as these significantly influence how overlapping is perceived \cite{stivers2009universals, clancy1996conversational}. In some cultures, conversational overlap is considered a sign of active engagement and is viewed positively. Users from these backgrounds may appreciate chatbot's overlap as a natural part of the interaction. Conversely, in cultures that prioritize clear turn-taking, such interruptions could be seen as rude or disruptive. This cultural variability underscores the need for configuration to be adaptable. By learning and adjusting to the conversational norms of individual users over time, the AI chatbot can better align its behavior with the user's cultural background.

\section{Limitations and Conclusion}\label{limitations}
One limitation of our study is that the majority of participants were native Korean speakers, with some also speaking Nepali and Indonesian as their primary languages. Although we assessed participants' English fluency during recruitment, conducting the user study sessions in English may have affected the interaction results. 

Additionally, cultural dynamics in conversational patterns might have affected the interactions, which could have impacted our findings but were not explicitly considered. Future research could explore how the chat interface performs with participants from diverse linguistic backgrounds.

Furthermore, our participant pool lacked age diversity, as most of the 34 participants were in their twenties. As discussed, OverlapBot's ability to create a sense of presence and companionship could lead to different interaction dynamics depending on the age, which we did not explore in this study. This limitation may have influenced our results, and further research is needed to understand how OverlapBot performs in interactions with a broader age range, particularly with older individuals \cite{DBLP:conf/chitaly/ValtolinaH21, DBLP:conf/icaai/KimKR22}. 

There is a technical issue when users begin typing a response, delete it, and then input a different prompt. In such cases, the chatbot might prematurely generate a response based on the initial input, which may no longer be relevant to the final prompt. This delay in response arises from the limitations in handling multi-threading when generating responses through the Huggingface API. Therefore, we chose to rely on generating responses on a single thread, which sometimes led to delays in response times. This was also perceived by users and may be solved by a specialized decoding algorithm for LLMs. 

Despite its limitations, we believe the introduction of OverlapBot to explore overlap in human-LLM text-based interaction is a worthwhile effort as an initial step. In this paper, we investigated the potential for integrating overlapping behaviors into text-based human-LLM communication, drawing parallels to human-to-human conversations, where overlap plays a key role in creating smooth exchanges. Our formative study with a research probe revealed that people instinctively use overlapping, such as preemptive answering and backchanneling, even in text-based interactions. Building on these insights, we developed OverlapBot, an AI chat prototype designed to replicate these behaviors. Our user study demonstrated that users found text-based overlap in human-LLM interactions to be more natural and efficient than the traditional turn-taking chat systems. Through this work, we identified new interaction patterns, explored the design space for overlapping interactions, and provided design guidelines for implementing these features in AI systems. We found that when allowed to overlap, people instinctively embrace it, revealing new possibilites for chatbot design.

\bibliographystyle{ACM-Reference-Format}
\bibliography{sample-base}

\appendix

\section{AI Chatbot}\label{sec:model-building}
To equip the chatbot with the overlapping capabilities, we addressed the technical challenges by finetuning an LLM with publicly available datasets engineered to our purpose. We will describe the challenges we encountered, the process we used to solve them, and the resulting outcomes.

\subsection{Technical Challenge}
During our usability evaluation in the design process, where we addressed issues iteratively, we initially built OverlapBot using the ChatGPT API. We experimented with various prompts and instructions to maintain a smooth conversation flow with overlapping. However, we found that its performance was inadequate for generating natural and interactive conversations. A significant issue was that ChatGPT overlapped with the user's typing too frequently, disrupting the conversation flow and causing friction. For example, when a user typed \textit{``how about doing someth,''} ChatGPT might prematurely interject with \textit{``..ing fun?''} This behavior revealed the model's difficulty in managing overlap appropriately. To resolve this challenge and equip the chatbot with better overlapping capabilities, we opted to finetune an open-source LLM with publicly available data engineered to our purpose, specifically refining its ability to handle overlapping interaction patterns. 

\subsection{Finetuning Strategy and Evaluation}

We selected the Llama3-8B model as the base for finetuning our chatbot. Using parameter-efficient finetuning \cite{DBLP:conf/iclr/HuSWALWWC22} through the Huggingface API, we fetched the model parameters. The chatbot was trained to generate the next conversational utterance based on the provided conversation context. 

To effectively evaluate handling of overlapping messaging, we established three key metrics. The first metric assesses chatbot’s ability to time its overlapping appropriately - Timing Classification. For instance, if a user is in the middle of typing \textit{``have you painted,''} the chatbot must decide whether it is more suitable to overlap with user's typing at that moment or to wait. The chatbot faces two options here: \textit{[Overlap]} or \textit{[Await]}. The second metric evaluates the chatbot’s ability to generate the appropriate dialogue actions if it decides to overlap rather than wait - Dialogue Act Classification. Continuing the example, if the chatbot chooses to overlap with user's typing, it must then decide whether to produce an understanding reaction, like \textit{``[Understanding],''} or to initiate an answer, indicated by \textit{``[Answer].''} The third metric focuses on the appropriateness of the utterance generated following the choice made in the second phase - Utterance Generation. For instance, if the chatbot selects \textit{[Understanding]}, it should generate a supportive reaction like \textit{``yeah''} or \textit{``keep going.''} If it selects \textit{[Answer]}, it should respond with something like \textit{``Yeah, I painted something recently!''} We trained Llama3-8B based on this logic ([Await/Overlap] [If Overlap==Dialogue Act] Utterance). To evaluate these abilities, we used F1 scores for the first two metrics and Bleu/Rouge scores for the third. 

\begin{table}[t!]
\caption{OverlapBot Evaluation Results}
\label{tab:automatic-result}
\resizebox{\columnwidth}{!}{%
    \begin{tabular}{l|rrrr|rrrr|rr}
    \toprule
               & \multicolumn{4}{c|}{Timing} & \multicolumn{4}{c|}{Dialogue Acts} & \multicolumn{2}{c}{Utterance} \\
    Models    & Acc. & Prec. & Rec. & F1  & Acc. & Prec. & Rec. & F1  & Bleu & Rouge-L \\
    \midrule
    Llama3 8B  & 0.46 & 0.50 & 0.50 & 0.46 & 0.59 & 0.79 & 0.50 & 0.37 & 0.27 & 0.11 \\
    GPT4o      & 0.50 & 0.47 & 0.47 & 0.47 & 0.73 & 0.74 & 0.76 & 0.73 & 0.08 & 0.12 \\
    GPT4 turbo & 0.50 & 0.46 & 0.46 & 0.46 & 0.75 & 0.77 & 0.77 & 0.73 & 0.26 & 0.12 \\
    \textbf{OverlapBot} & \textbf{0.67} & \textbf{0.65} & \textbf{0.65} & \textbf{0.65} & \textbf{0.85} & \textbf{0.87} & \textbf{0.78} & \textbf{0.80} & \textbf{0.48} & \textbf{0.30} \\ 
    \bottomrule
    \end{tabular}%
}
\end{table}

\begin{figure*}
    \centering
    \includegraphics[width=\textwidth]{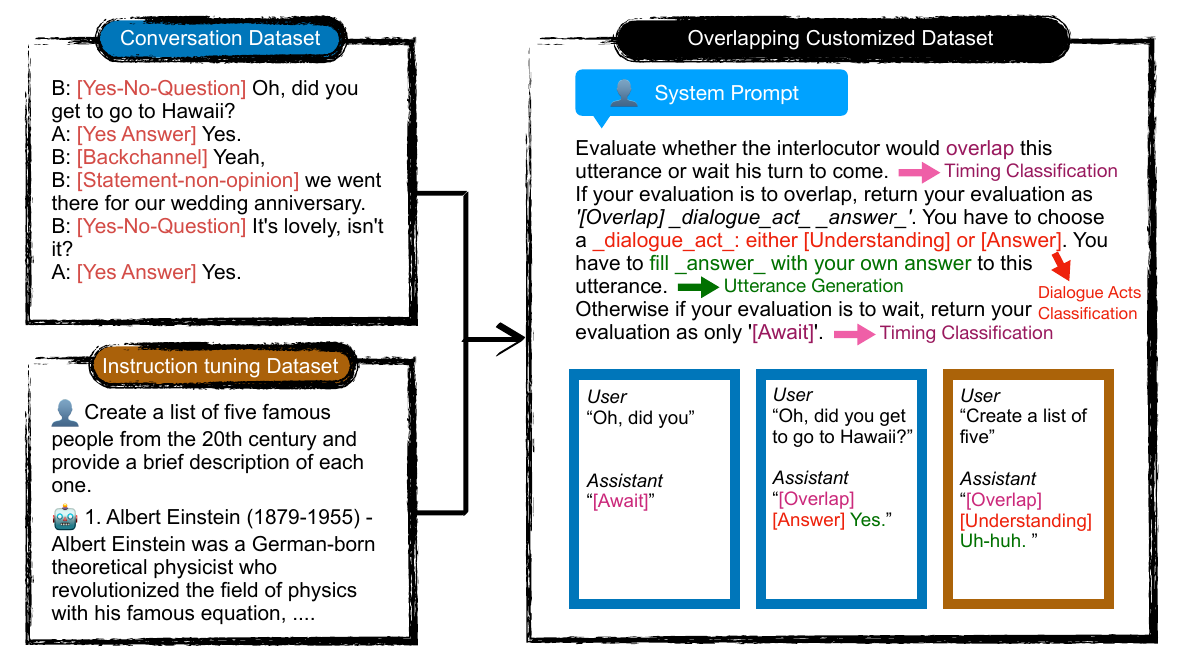}
    \caption{An example of a customized dataset engineered for modeling overlapping behaviors in human-AI interactions.}
    \label{fig:datamanipulation}
\end{figure*}  

\subsection{Data Manipulation}
\label{Appendix-datamanipulation}
We utilized two distinct datasets including conversation dataset and instruction-tuning dataset. We manipulated both datasets by inserting tags for timing classification ([Overlap]/[Await]), dialogue act classification ([Understanding]/[Answering]), and utterance generation. An example of customized dataset for overlapping is shown in Figure \ref{fig:datamanipulation}. 

The first dataset is the Switchboard Dialogue Act Corpus (SWDA), which consists of 1,155 five-minute telephone conversations between pairs of participants \cite{godfrey1992switchboard}. In these conversations, 440 speakers engage in discussions on various topics, such as child care, recycling, and news media, resulting in a total of 221,616 utterances. We chose the SWDA due to its comprehensive dialogue act annotations, which include 43 different dialogue actions, among them overlapping signals like backchanneling and sentence completion. For our purposes, we preprocessed the SWDA by consolidating all overlapping signals into a single [Understanding] tag. This simplification was necessary because the dataset includes numerous tags that are infrequently represented (less than 1\%), making it challenging for the chatbot to effectively learn from them. By reducing the complexity of the classification problem, we aimed to improve the chatbot's performance. 

The second dataset we used was an instruction-tuning dataset \footnote{\url{https://huggingface.co/datasets/yahma/alpaca-cleaned}}. Given that the SWDA is primarily a conversational dataset, we recognized that a model finetuned solely on SWDA might struggle with task-oriented dialogues. At the same time, to better align instruction-tuning dataset with the requirements of overlapping tasks, we manipulated the dataset by inserting [Overlap] or [Await] tags as responses. For instance, in an instruction like \textit{``Please write me an essay about a random topic,''} we might edit it to \textit{``Please write me an''} and then insert a response such as \textit{``[Await]''} or \textit{``[Overlap] [Understanding] Yeah.''} The understanding utterances were randomly selected from a set of predetermined choices based on the backchanneling utterances in SWDA. This adjustment was intended to enhance the chatbot's ability to handle overlap in more structured, task-oriented interactions.

\subsection{Results}
 
As seen in Table \ref{tab:automatic-result}, chatbot demonstrated superior performance across all evaluated tasks, including timing classification, dialogue act classification, and utterance generation. During evaluation, we gave same prompts and evaluation datasets to all chatbots. Based on the system prompt and given user utterance, chatbots generated responses. We evaluated the chatbots' automatic performance on classification accuracy (accuracy, precision, recall, F1 score) and reference-based generation accuracy (Bleu \cite{papineni2002bleu}, Rouge-L scores \cite{lin2004rouge}). These results indicate that the finetuning process effectively enhanced the chatbot's overlapping ability in a way that outperforms both the baseline chatbots.

\end{document}